\documentclass[aps,twocolumn,floats,prl,nofootinbib,10pt,longbibliography,superscriptaddress]{revtex4-1}

\usepackage{comment}
\usepackage[dvips]{graphicx} %
\usepackage{graphicx,amsmath,amsfonts,amssymb,slashed,float,hyperref}
\usepackage[normalem]{ulem}
\usepackage{bbold,wasysym}
\usepackage{graphicx}
\usepackage{array,multirow}
\usepackage[utf8]{inputenc}
\usepackage{scalerel}
\usepackage{cleveref}

\usepackage[usenames,dvipsnames]{xcolor} 

\usepackage{soul}
\usepackage{bm}

\definecolor{RoyalBlue}{rgb}{0.25,.41,.88}
\definecolor{celestialblue}{rgb}{0.29, 0.59, 0.82}

\setstcolor{Blue}

\def\AB#1{\textcolor{Magenta}{AB}}

\newcommand{\be}{\begin{equation}}
\newcommand{\ee}{\end{equation}}
\newcommand{\bea}{\begin{eqnarray}}
\newcommand{\eea}{\end{eqnarray}}
\newcommand{\Beq}{\begin{equation}\begin{aligned}}
\newcommand{\Eeq}{\end{aligned}\end{equation}}

\definecolor{cerulean}{rgb}{0., 0.62,0.7}

\newcommand{\Mpl}{M_{\rm pl}}

\usepackage{color}
\usepackage{ifthen}
\newboolean{editorial}
\setboolean{editorial}{true}
\newcommand{\editorial}[2]{\ifthenelse{\boolean{editorial}}{\textcolor{red}{[\textsf{\textbf{{#1}}}: }\textcolor{blue}{\textsf{{#2}}}\textcolor{red}{]}}{}}

\usepackage{xcolor}

\begin{document}

\title{Kinetic Preheating after $\alpha$-attractor Inflation}

\author{Peter Adshead}
\affiliation{Illinois Center for Advanced Studies of the Universe \& Department of Physics, University of Illinois at Urbana-Champaign, Urbana, Illinois 61801, U.S.A.}

\author{John T. Giblin, Jr.}
\affiliation{Department of Physics, Kenyon College, Gambier, Ohio 43022, U.S.A.}
\affiliation{Department of Physics, Case Western Reserve University, Cleveland, Ohio 44106, U.S.A.}
\affiliation{Center for Cosmology and AstroParticle Physics (CCAPP) and Department of Physics, Ohio State University, Columbus, Ohio 43210, U.S.A.}

\author{Reid Pfaltzgraff-Carlson}
\affiliation{Department of Physics, Kenyon College, Gambier, Ohio 43022, U.S.A.}

\begin{abstract}
    We study preheating via kinetic couplings after dilaton-axion $\alpha$-attractor inflation. We focus on E-model $\alpha$-attractor driven inflation where the inflaton is kinetically coupled to an ultralight axion. In this class of models, the kinetic coupling is related to the form of the potential, and once the  amplitude of the scalar curvature spectrum as well as the tensor-to-scalar ratio are specified, the model has no free parameters. We find that kinetic preheating can be extremely efficient, with stronger preheating occurring at parameter values corresponding to smaller values of the tensor-to-scalar ratio. Preheating becomes extremely efficient below $r \lesssim 1.6\times 10^{-5}$.
   
\end{abstract}

\maketitle

\section{Introduction}

 Non-minimal $\alpha$-attractor classes of inflationary models present a novel opportunity to study preheating in a scenario where the inflaton's potential and kinetic couplings to other fields share a single parameter.   These kinetically coupled fields occur generically in $\alpha$-attractor and supergravity models \cite{Kallosh:2013maa, Linde:2018hmx, Braglia:2020eai, Kallosh:2022vha}, as well as in various string constructions, such as Fibre Inflation \cite{Cicoli:2008gp}, and generalized Einstein theories \cite{Starobinsky:2001xq, DiMarco:2002eb}. In the simplest realizations of $\alpha$-attractors \cite{Kallosh:2013maa},  kinetic couplings play only a small role during preheating due to the strong attractor nature of the background multifield trajectories \cite{Iarygina:2018kee}. These trajectories generically end inflation along directions where the kinetic couplings vanish during the early stages of preheating. Nonetheless, preheating after $\alpha$-attractor inflation can be very efficient \cite{Krajewski:2018moi, Iarygina:2018kee, Iarygina:2020dwe}, leading to significant gravitational wave backgrounds \cite{Li:2020qnk, Krajewski:2022ezo}

In this work, we consider scenarios in which the inflaton enters the reheating phase with a significant kinetic coupling to an additional scalar degree of freedom. We specialize to exponential type couplings of the sort that arise in supergravity constructions such as multifield generalizations of $\alpha$-attractors \cite{Kallosh:2013daa, Achucarro:2017ing}. An explicit example of inflationary model where inflation can end with a significant kinetic coupling is Hypernatural inflation \cite{Linde:2018hmx, Kallosh:2022feu}, where the inflaton can be either the dilaton or axion, or some combination of both. These dilaton-axion models have been used to construct two-stage inflationary models that generate large amplitude scalar fluctuations near the end of inflation to seed primordial black holes and induce large-amplitude gravitational waves \cite{Braglia:2020eai, Kallosh:2022vha}. 

We consider inflation to proceed along the dilatonic direction on an E-model $\alpha$-attractor potential. After fixing the amplitude of the fluctuations to match those observed in the cosmic microwave background, the resulting model has only one free parameter, the scale $\mu$ which parameterizes the curvature of the K\"{a}hler manifold. This parameter sets the tensor-to-scalar ratio, $r$, and also controls the strength of the kinetic coupling. We explore the efficiency of preheating as a function of $\mu$, and find that preheating into the spectator field can be extremely efficient for strongly curved manifolds, which correspond to low values of the tensor-to-scalar ratio $r < 1.6\times 10^{-5}$ in these models.

\begin{figure*}
\centering
\includegraphics[width=\linewidth]{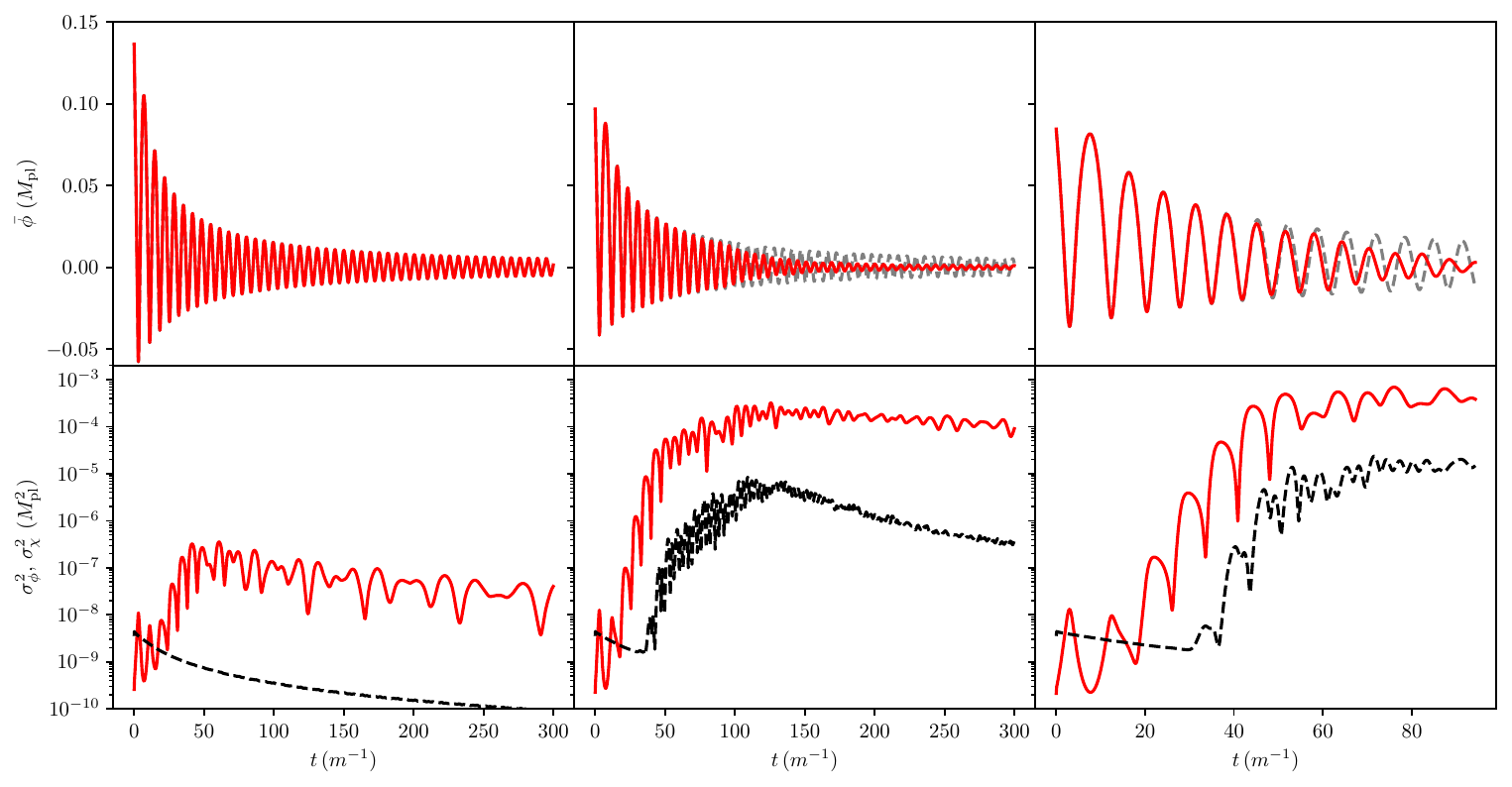}[h!]
\caption{\label{fig:meansvariancecomp}Statistics of preheating for $\mu = 0.104 \,\Mpl$ (left panels), $\mu = 0.0705 \,\Mpl$ (middle panels) and $\mu = 0.0602 \,\Mpl$ (right panels).  The top panels show the average value of the field in each simulation (red) as well as the solution $\bar{\phi}$ from the homogeneous limit of \cref{eq:eomphi} (gray, dashed).  The bottom panels show the variance of the inflation, $\sigma_\phi^2$ (red), alongside the variance of the axion, $\sigma_\chi^2$ (black, dashed).  Note that the timescale for the $\mu = 0.0602 \,\Mpl$ simulation is shorter than the other two cases.}
\end{figure*}

\begin{figure*}
\centering
\includegraphics[width=\linewidth]{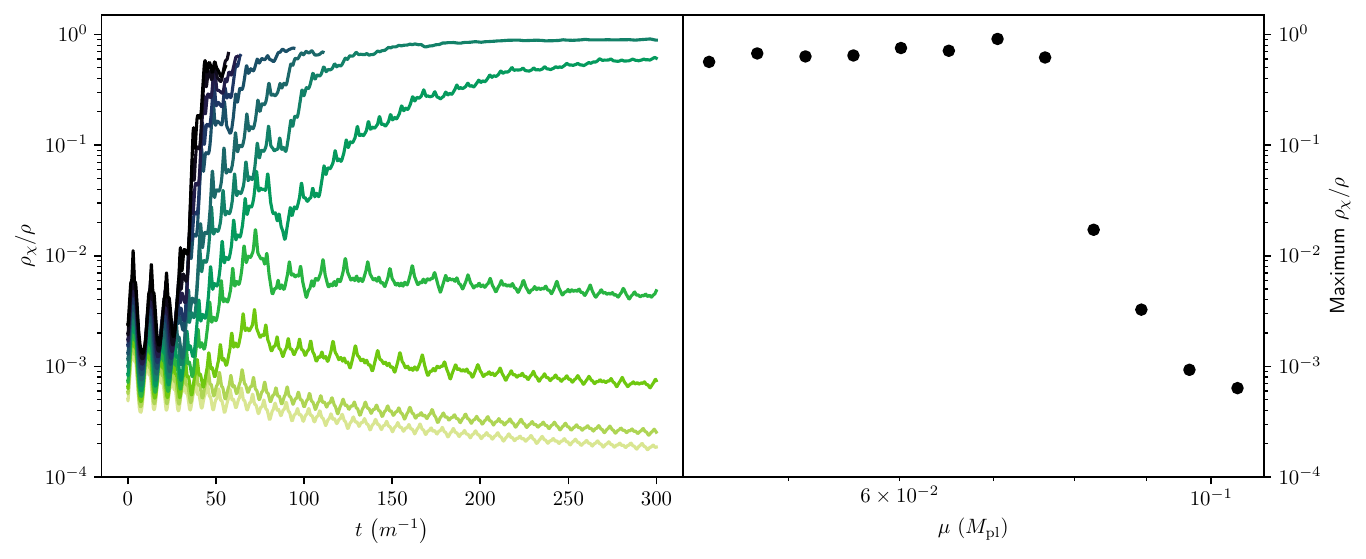}
\caption{\label{fig:ratiosforM}The ratio of energy deposited into the $\chi$ field as a function of $\mu$.  The left panel shows the fraction of energy in the $\chi$ field as a function of the time in the simulation, $\bar{\rho}_\chi/\bar{\rho}$; darker colors (top) correspond to the data produced by simulations with a lower values of $\mu$ ranging from $\mu = 0.0439\,\Mpl$ (upper leftmost curve) to $\mu = 0.104\,\Mpl$ (lowest curve).  The right panel shows the maximum value of the ratio of $\bar{\rho}_\chi/\bar{\rho}$ for each value of $\mu$ evaluated after preheating has begun, $t \gtrsim 45\,m^{-1}$.}
\end{figure*}

We use natural units, which set $\hbar = c = 1$, and define the reduced Planck mass
$\Mpl = 1/\sqrt{8\pi G}$. We use the ``mostly plus"  metric convention for a homogeneously expanding Friedmann-Lema\^itre-Robertson-Walker background and repeated/contracted Greek spacetime indices are summed via the Einstein summation convention.  Over-bars, e.g. $\bar{\phi}$, as well as angled-brackets,  $\left<\phi\right>$, denote constant-time grid-averaged quantities.


\section{The model and background}

We consider a dilaton-axion model of inflation with dilatonic scalar field, $\phi$, and an axion field, $\chi$,
\begin{align}
\mathcal{L} = -\frac{M_{\rm Pl}^2}{2}R -\frac{1}{2}\left(\partial\phi\right)^{2} - \frac{W(\phi)}{2}(\partial\chi)^2 - V(\phi) ,
    \label{eq:action}
\end{align}
where $R$ is the Ricci scalar.    The inflaton, $\phi$, is subject to an E-model $\alpha$-attractor potential \cite{Kallosh:2013maa, Linde:2018hmx}
\begin{equation}\label{eqn:Emodelpotential}
V(\phi) = \frac{m^2\mu^2}{2}\left(1-e^{-\phi/\mu}\right)^2,
\end{equation}
where $\mu$ is a free parameter. The height of the potential, parametrized by $m$, however, is set by matching to the amplitude of the scalar spectrum as measured by Planck \cite{Planck:2018jri}
\begin{equation}
\frac{H_{\rm 50}^2}{8\pi \Mpl^2 \epsilon_{\rm 50}} = 2\times 10^{-9},
\label{eq:setmfromepsilon}
\end{equation}
where $H_{\rm 50}$ and $\epsilon_{\rm 50}$ are the Hubble rate, $H = \dot{a}/a$, and slow-roll parameter, $\epsilon = -\dot{H}/H^2$, evaluated 50 $e$-foldings before the end of inflation.  In these models  the tensor-scalar ratio, $r$, directly depends on $\mu$, while the scalar tilt is approximately independent \cite{Kallosh:2013hoa},
\begin{equation}
r = \mu^2\frac{8}{N^2}\, ,\quad n_s = 1-\frac{2}{N},
\label{eq:rfromeq}
\end{equation}
where $N$ is the number of $e$-foldings before the end of inflation. While \cref{eq:rfromeq} is an approximation, numerical solutions to the homogenous evolution of \cref{eq:setmfromepsilon} evaluated $N = 50$ $e$-foldings before the end of inflation, and using the standard $r = 16 \epsilon_{50}$ agree with \cref{eq:rfromeq} within ten percent.

The kinetic coupling between the inflation and the axion is not a free function in this model
 \cite{Braglia:2020eai, Kallosh:2022vha},
\begin{equation}
W(\phi) = e^{2\phi/\mu},    
\label{coupling term}
\end{equation}
thereby leaving the entire inflaton-axion system parameterized by $\mu$ alone.  This choice of potential is not unique \cite{Kallosh:2022feu}, but it provides an intriguing opportunity to study a preheating system in the absence of free parameters. For this work, we assume that the field $\chi$ has no vacuum expectation value during inflation.\footnote{Note that this does not result in geometric destabilization of the ultralight $\chi$ field during inflation \cite{Cicoli:2018ccr}, as shown in Ref.\ \cite{Achucarro:2017ing}} 

In the absence of a coupled axion, {\sl oscillons} \cite{Bogolyubsky:1976nx,Bogolyubsky:1976sc,Gleiser:1993pt,Copeland:1995fq,Kasuya:2002zs,Saffin:2006yk,Hertzberg:2010yz,Amin:2011hj,Salmi:2012ta,Gleiser:2019rvw,Antusch:2019qrr,Ibe:2019vyo,Zhang:2020bec,vanDissel:2023zva} can form from these asymmetric potentials \cite{Antusch:2017flz,Hasegawa:2017iay}, which can also produce gravitational waves \cite{Antusch:2016con,Amin:2018xfe}.  While these self-resonances also occur in the model presented here, the additional instabilities in the axion field can dominate the preheating phase.

The model in eq.\ \eqref{eq:action} leads to the classical equations of motion for the fields $\phi$ and $\chi$
\begin{align}
 \ddot{\phi} +3H \dot{\phi} - \frac{\nabla^2 \phi}{a^2} + \frac{\partial V}{\partial \phi} = & \frac{e^{2\phi/\mu}}{\mu}\left(\dot{\chi}^2 -\frac{1}{a^2} \vec{\nabla}\chi \cdot \vec{\nabla}\chi\right) ,
   \label{eq:eomphi} \\
\ddot{\chi} + 3H\dot{\chi} - \frac{\nabla^2 \chi}{a^2} = & 
- \frac{2}{\mu}\left(\dot{\chi}\dot{\phi} -\frac{1}{a^2} \vec{\nabla}\chi \cdot \vec{\nabla}\phi\right)\! .
\label{eq:eomchi}
\end{align}
The expansion of the universe is taken to be rigid and the evolution of the scale factor is self-consistently calculated from the Friedmann equation, $3\Mpl^2 H^2 = \rho$. The energy density receives contributions from both the inflaton and axion, $\rho = \rho_\phi + \rho_\chi$, where
\begin{align}
\rho_\phi &= \dot{\phi}^2 + \frac{1}{a^2}\vec{\nabla}\phi \cdot \vec{\nabla}\phi + V(\phi)\, ,
\label{eq:rhophi}\\
\rho_\chi &= e^{2\frac{\phi}{\mu}}\left(\dot{\chi}^2 +\frac{1}{a^2} \vec{\nabla}\chi \cdot \vec{\nabla}\chi\right). 
\label{eq:rha}
\end{align}
To study the early stages of kinetic preheating, we ignore inflaton fluctuations, and linearize the equation of motion for the canonically normalized axion field $\varphi = \sqrt{a^3W(\phi)}\chi$. In Fourier space, this reads
\begin{align}\label{eqn:fouriereom}
\ddot{\varphi}_k + \left(\frac{k^2}{a^2}-\left(\frac{3}{2} H +\frac{ \dot{\phi}}{\mu}\right)^2-  \left(\frac{3}{2} \dot{H} +\frac{ \ddot{\phi}}{\mu} \right)\right)\varphi_k = & 0.
\end{align}
A tachyonic instability exists when the effective mass of $\varphi$ is negative.  We then estimate that significant instabilities exist during reheating for momenta in the band
\begin{align}\label{eqn:instabband}
H < \frac{k}{a} <  \frac{\dot{\phi}}{\mu}, \sqrt{\frac{\ddot{\phi}}{\mu}},
\end{align}
assuming $H  \ll \dot{\phi}/\mu$.  Because $m$ controls the dynamics of $\phi$ in this regime, $\dot{\phi} \sim m\phi$, while $H^{-1} \sim 50 m^{-1}$ (see below), wide instability bands can exist even when $\phi_0/\mu \sim 1$,  $\phi_0$ is the value of the inflaton at the end of inflation. 


\section{Initial Conditions and Numerical Parameters}

To set our initial conditions we first numerically integrate the homogeneous mode of the inflation field during inflation.  We simulate many $e$-foldings of inflation for each value of $\mu$, thereby ensuring that we are on the inflationary attractor $50$ $e$-foldings before the end of inflation.  We use these to calculate $m$ from \cref{eq:setmfromepsilon} as well as the homogenous values $\phi_0 = \bar{\phi}$ and $\dot{\phi}_0 = \bar{\dot{\phi}}$ at the end of inflation.  

We employ {\sc Gabe} \cite{Child:2013ria} to conduct nonlinear simulations during the preheating period.  For the values of $\mu$ that we study, $H^{-1}\sim 50\,m^{-1}$ at the end of inflation, although this number varies from $28\,m^{-1}$ for the largest value of $\mu$ to $59\,m^{-1}$ for the smallest value of $\mu$.  We give $\phi$ and $\chi$ Bunch-Davies initial conditions,
\begin{equation}
 \left<\left|{\phi(k)}\right|^2\right> = \left(2a\omega_\phi \right)^{-1}
 \end{equation}
 and 
\begin{equation}
\left<\left|\chi(k)\right|^2\right> = \left(2a\omega_\chi W(\bar{\phi})\right)^{-1}.
\end{equation}
The extra factor of $W$ in the $\chi$ initial conditions represents the fact that $\varphi$ is the canonically quantized field.  The frequencies, $\omega = \sqrt{k^2+m_{\rm eff}^2}$, are set by the wave number and the curvature of the potential at the end of inflation, $m^2_{\rm eff} = \partial^2 V/\partial \phi^2$ for the inflaton, and $m_{\rm eff}^2 = 0$ for the axion.\footnote{Note that this is not strictly true as can be seen from eq.\ \eqref{eqn:fouriereom}. However, it remains an excellent approximation until very close to the end of inflation, where the effective mass quickly becomes large and negative. Our initial conditions therefore likely underestimate the subsequent amplification of $\chi$ during preheating.}  Since the effective mass is slightly negative at the end of inflation, $m^2_{\rm eff} = - 0.125 \,m^{2}$, we set $L_0 = 15\,m^{-1}$ which ensures that the smallest non-zero mode of the box $k_{\rm min} = 2\pi/L_0$ is not tachyonic when the simulation begins.

For all of the simulations here the grid has $N^3 = 256^3$ points with a time-step of $\Delta t = \Delta x / 30 = L_0/N/30$.  Inhomogeneities in the fields can be parameterized by the variance, $\sigma^2$, e.g. 
\begin{equation}
\sigma^2_{\phi} = \left<\phi^2 - \bar{\phi}^2\right>.
\end{equation}


\section{Results}

Preheating is characterized by the explosive production of particles due to the  periodic dynamics of the homogeneous inflaton background \cite{Traschen:1990sw, Shtanov:1994ce,Kofman:1994rk, Kofman:1997yn}. In momentum space, this results in the parametric or tachyonic amplification of axion momenta that fall into unstable regions of the relevant Floquet chart. In the case of nonlinear self-interactions induced by anharmonicity of the potential in \cref{eqn:Emodelpotential}, resonance results in amplification of the inflaton field itself.  When preheating is efficient and large amplitude fluctuations are excited, a second phase can occur in which backscattering excites further field inhomogeneities which can lead to the decay of the homogeneous mode of the inflaton. For the kinetic coupling we consider in this work, preheating is dominated by energy transfer to the axion.  

The value of $\mu$ has two effects on the dynamics. First it changes the hierarchy between $m$ and $H$ at the end of inflation, and secondly it directly controls the strength of the kinetic coupling and therefore the width of the instability bands in eq.\ \eqref{eqn:instabband}. We therefore expect stronger preheating effects at smaller values of $\mu$.  For larger values of $\mu$, we observe weak preheating.  This can be seen in the left panels of \cref{fig:meansvariancecomp}, for the case where $\mu = 0.104 \,\Mpl$.  In these weak cases, there is some resonance--characterized by an increase in the variance of the axion, $\chi$, with little back-scattering.  In these cases the average of $\phi$ does not fragment and matches the numerical evolution of the homogeneous limit of \cref{eq:eomphi}.  The middle panels of \cref{fig:meansvariancecomp} show a marginal case, $\mu = 0.0705 \,\Mpl$, where the modes of $\chi$ are amplified, followed by back-scattering onto $\phi$; the zero mode of the inflation also decays from its homogeneous limit.  For values of $\mu \lesssim 0.07 \,\Mpl$, the preheating process appears quite efficient, as in the right panels of \cref{fig:meansvariancecomp}, with the variance of $\chi$ growing within a few oscillations of the homogenous mode.  In these cases, our simulations terminate due to significant scattering into higher-$k$ modes.

In \cref{fig:ratiosforM} we show the results of our simulations over a wide range in $\mu$.  We probe a range of $\mu$ values, sampled logarithmically, from $\mu = 0.0439\,\Mpl$ to $\mu = 0.104\,\Mpl$.  For values of of $\mu$ below the threshold of $\mu \approx 0.8 \,\Mpl$, we see that preheating generically arises in these models. As $\mu$ increasing, the preheating efficiency quickly decreases. For the case of $\mu=0.0705\,\Mpl$, which we identify as the borderline case between incomplete and complete preheating, this gives $r\approx1.6\times 10^{-5}$.  In cases where $r$ is less than this value, we expect that the kinetic coupling to be the dominant reheating channel.  

In models where no kinetic coupling exists, oscillons are formed for $\mu \lesssim 0.08$ \cite{Aurrekoetxea:2023jwd}.  Incidentally, kinetic preheating becomes efficient near this same threshold, implying that the kinetic channel dominates over self-resonance and suppresses oscillon production.

\section{Discussion and conclusions}

In this work, we have studied preheating in a class of kinetically coupled inflationary models. These types of kinetic couplings arise naturally in classes of non-minimal $\alpha$-attractor inflation currently favored by observational data. Specifically, we have considered a dilatonic inflaton coupled via an exponential kinetic coupling to an ultralight axion field. For observationally-relevant values of $\mu$, strong preheating arises without the need for any tuning. We find preheating is stronger at lower values of the tensor-to-scalar ratio, with the transition to efficient preheating occurring for $r\approx1.6\times 10^{-5}$.

Finally, we anticipate that, analogously to the kinetically-coupled gauge field case  \cite{Deskins:2013dwa, Adshead:2017xll}, gravitational  wave production will be extremely efficient in these scenarios \cite{Adshead:2018doq}, and may lead to constraints on the scenario through cosmic microwave background measurements of the effective number of relativistic species, $N_{\rm eff}$ \cite{Adshead:2019lbr, Adshead:2019igv}. Further, in the region of parameter space where preheating is extremely efficient, we expect large density inhomogeneities  to be produced, potentially collapsing to form compact structures such as black holes \cite{Adshead:2023mvt}. These, and other investigations will be published separately \cite{Adshead:2023_1, Adshead:2023_2}.


\acknowledgments
We thank Mustafa Amin and Zachary Weiner for early conversations regarding kinetic couplings.  P.A. and J.T.G. gratefully acknowledge support from the Simons Center for Geometry and Physics, Stony Brook University at which some of the research for this paper was performed. J.T.G. \ is also grateful for the hospitality of the Illinois Center for Advanced Studies of the Universe at the University of Illinois at which much of this work was conducted. P.A.\ is supported in part by the United States Department of Energy, DE-SC0015655. 
J.T.G. and R.P.\ are supported in part by the National Science Foundation, PHY-2309919.  The simulations presented here were carried out at the Ohio Super Computer Center \cite{OhioSupercomputerCenter1987}.

\bibliography{KineticPreheat}

\end{document}